\documentclass[9pt,twocolumn,twoside]{opticajnl}
\journal{opticajournal} 

\setboolean{shortarticle}{true}

\usepackage{lineno}

\title{Quantum entanglement distribution coexisting with high-rate, broadband classical optical communications over a real-world fiber connecting remote, synchronized nodes}

\author[1,2,*]{Gina M. Talcott}
\author[1]{Ahnnika I. Hess}
\author[1,3]{Laura d'Avossa}
\author[4]{Scott J. Kohlert}
\author[5]{Fei I. Yeh}
\author[5]{Jim Hao Chen}
\author[5]{Joe J. Mambretti}
\author[6]{Tim M. Rambo}
\author[1,7]{Gregory S. Kanter}
\author[1]{Jordan M. Thomas}
\author[1,2,**]{Prem Kumar}

\affil[1]{Center for Photonic Communication and Computing, Department of Electrical and Computer Engineering, Northwestern University, 2145 Sheridan Road, Evanston, IL 60208, USA}
\affil[2]{Graduate Program in Applied Physics, Northwestern University, Evanston, IL 60208, USA}
\affil[3]{University of Naples Federico II, Via Claudio 21, 80125 Napoli NA, Italy}
\affil[4]{Ciena Corporation, 7035 Ridge Road, Hanover, MD 21076, USA}
\affil[5]{International Center for Advanced Internet Research, Northwestern University, 750 N. Lake Shore Drive, Chicago, IL 60611, USA}
\affil[6]{Quantum Opus LLC, 14841 Keel Street, Plymouth, MI 48170, USA}
\affil[7]{NuCrypt LLC, 1460 Renaissance Drive Suite \#205, Park Ridge, IL 60068, USA}

\affil[*]{ginatalcott@u.northwestern.edu}
\affil[**]{kumarp@northwestern.edu}

\begin{abstract} 
Compatibility with existing classical network infrastructure offers a scalable path towards deploying large-scale quantum networks. Here, we demonstrate O-band polarization-encoded quantum entanglement distribution over an installed 24.4\nobreakdash-km fiber while coexisting with a state-of-the-art fully-loaded C\nobreakdash-band classical communications line system and a picosecond-level precision L-band synchronization signal. The classical system carries two 800\nobreakdash-Gbps channels while the remainder of the C-band is filled with amplified spontaneous emission, as is standard for such state-of-the-art communications systems. We examine the spontaneous Raman scattering spectrum generated from this broadband C\nobreakdash-band light and offer insights into wavelength allocation for O\nobreakdash-band quantum channels. Optimal wavelength selection and narrow filtering enable well-preserved Bell state fidelity when coexisting with 21.4\nobreakdash-dBm aggregate launch power across the C\nobreakdash-band suitable for 36\nobreakdash-Tbps transmission. To the best of our knowledge, this is the first implementation of entanglement-based quantum communications between two remote nodes coexisting with independent classical communications traffic. We demonstrate coexistence of quantum entanglement with ultra-high power levels and record classical bandwidth, offering promise for real-world entanglement-based networking integrated within high-capacity communications infrastructure.
\end{abstract}
\setboolean{displaycopyright}{false}
\begin{document}
\maketitle
\vspace{-18pt}
\section{Introduction}
\vspace{-2pt}
Recent years have seen rapid progress in quantum networking research, a topic of current importance for distributed quantum computing, quantum cryptography, sensing, and metrology~\cite{wehner_review,kimble_quantum_internet}. As researchers work to develop quantum networks beyond laboratory research, it is important to consider how they will scale for future commercial or public use. Optical transmission at telecommunications wavelengths in fiber has a long history of successful use in intra- and inter-city networks for classical communications to enable global connectivity. This has motivated decades of work and significant advances in quantum communications over quantum-dedicated dark fiber~\cite{8node_entdist,ornl_network,valivarthi_tele2016,neumann_dark_entdist2022,du_cmos_entdist2025}. However, high demand for optical fiber and high cost of installing new fiber links suggest that integrating quantum networks into the existing classical communications infrastructure presents a more efficient and scalable path to developing widespread quantum networks. Wavelength-division multiplexing offers compatibility with dominant commercial optical networks and allows increased fiber capacity by housing many quantum and classical signals within the same fiber. Furthermore, "coexisting" classical and quantum networks can allow easy integration of classical signals for quantum network management (such as synchronization, routing information, or communicating measurement results)~\cite{quantum_wrapper, nist_sync, zhang_classical_decicisive}. 

As a significant step towards such integration, we demonstrate distribution of quantum entanglement in a real-world network infrastructure. Our quantum signal travels simultaneously with high-power, state-of-the-art classical C\nobreakdash-band communications in the same fiber, demonstrating coexistence with record bandwidth and ultra-high launch powers. Furthermore, we present the first (to our knowledge) real-world implementation of entanglement-based coexistence with independent classical communications. We distribute entanglement over an installed metropolitan fiber between synchronized quantum nodes in Evanston and Chicago, operating out of a telecommunications exchange facility. This reflects a transformative step forward in studies on coexistence, bringing hybrid quantum-classical networks out of the laboratory and towards fully deployable networks for academic, public, or commercial applications. 

Previous studies have demonstrated strong potential for hybrid quantum-classical networking in wavelength-division multiplexed (WDM) optical networks as early as 1997~\cite{townsend}, but the majority of experiments use weak coherent states (WCS) for discrete-variable quantum key distribution (QKD)~\cite{townsend, chapuran_qkd2009, aleksic_sprsqkd2015, dynes_qkd, eraerds_qkd2010, qi_qkd2010, patel_qkd2012, geng_qkd2021_ull, geng_qkd2021, gavignet_qkd, mao_qkd2018, honz_hollowcore_qkd2023, dou_qkd2024, beppu_qkd2025, peters_qkd2009, tanaka_qkd2008, grunenfelder_qkd2021, wang_qkd2017, dynes_cambridge_qkd, peng_qkd_osc2025}, continuous-variable QKD~\cite{kumar_qkd2015, kawakami_qkd2025, eriksson_qkd2019, hajomer_qkd}, or measurement-device-independent QKD~\cite{berrevoets_mdiqkd2022,valivarthi_mdiqkd2019} to distribute quantum information. A smaller number of works have demonstrated coexistence in research motivated for entanglement-based applications~\cite{hipp_qkd2016, fan_entdist_and_qkd2023, wang_entqkd2023, holloway_entqkd2011, luo_entdist_and_qkd2025,kapoor_fermi_argonne_sync, anirudh_argonne_sync, raju_fermi_sync, jordan_ent_dist, liang_ent_dist2006, sauge_entdist2007, yuan_entdist2019, jing_entdist2024, nist_ent_dist,mehmet_qwn_entdist2024, bearlinQ, jordan_tele}. Throughout this body of work, several wavelength allocation schemes have been investigated in order to minimize noise in the quantum links. Because spontaneous Raman scattering (SpRS) is typically the dominant noise contribution for WDM quantum networks, a large wavelength detuning between quantum channels and high-power classical channels can heavily reduce the negative impact of coexistence on quantum communications~\cite{jordan_ent_dist, wang_qkd2017}. The C\nobreakdash-band is generally preferred for transmission in fiber due to low propagation loss ($\sim0.2$\,dB/km), but several studies have investigated moving either the quantum or classical channels to the O\nobreakdash-band, at the cost of slightly increased propagation loss ($\sim0.3$\,dB/km). Despite this increased loss, these wide channel separations offer dramatic decreases in noise (orders of magnitude SpRS suppression in some cases) that well-compensate the extra loss, enabling robust quantum communications with coexisting classical signals~\cite{jordan_ent_dist}. Several works have utilized this to demonstrate successful entanglement-based communications for O\nobreakdash-band quantum/C\nobreakdash-band classical coexistence~\cite{hipp_qkd2016, bearlinQ, jordan_ent_dist, jordan_tele} and C\nobreakdash-band quantum/O\nobreakdash-band classical coexistence~\cite{kapoor_fermi_argonne_sync, nist_ent_dist, anirudh_argonne_sync, raju_fermi_sync}. This established research base provides strong proof-of-concept for quantum entanglement-based systems to coexist with WDM classical fiber networks, but more work is needed to demonstrate compatibility with real-world communications systems. 

Notably, many entanglement-based works demonstrate coexistence with only a low-power synchronization signal~\cite{sauge_entdist2007, nist_ent_dist, raju_fermi_sync, anirudh_argonne_sync} or a single-wavelength classical source with unidirectional~\cite{mehmet_qwn_entdist2024, holloway_entqkd2011, jing_entdist2024, liang_ent_dist2006, jordan_tele} or bidirectional~\cite{fan_entdist_and_qkd2023, luo_entdist_and_qkd2025, yuan_entdist2019, bearlinQ} propagation. Most of these works coexist with signals at launch powers between $-20$ to $0$\,dBm, which is significantly lower than the launch powers needed for high-capacity independent classical telecommunications traffic. Particularly, synchronization or polarization reference signals that aid quantum links can usually be attenuated to just above receiver sensitivity thresholds~\cite{nist_ent_dist}. While independent classical communications can also be attenuated, it is not typical in conventional operation and would limit data rates or require modification to existing physical layers. Furthermore, current state-of-the-art classical telecommunications systems often fill the entire C\nobreakdash-band (and sometimes L\nobreakdash-band) with strong classical light~\cite{C+L_classical}. As such, it is important to provide analysis of quantum systems coexisting with high-power, broadband classical communications to replicate sharing bandwidth on an active, real-world fiber. We also note that commercial systems are unlikely to favor allocation for quantum channels, instead prioritizing high data rate transmissions in the low-loss C\nobreakdash-band. Further, some high-bandwidth classical systems functionally prohibit quantum allocation within the same transmission bands by filling all available spectrum with amplified spontaneous emission (ASE) for channel monitoring~\cite{C+L_classical}. However, the O\nobreakdash-band is often left available in commercial long-haul networks, making it a prime candidate for developing quantum networks within the classical infrastructure. Despite the prominence of fully-loaded C\nobreakdash-band classical systems in state-of-the-art commercial communications, only a few works with WCS (and none using entanglement-based sources) have utilized a fully-loaded band for classical communications~\cite{wang_qkd2017, gavignet_qkd, honz_hollowcore_qkd2023, kawakami_qkd2025, dou_qkd2024, beppu_qkd2025, eriksson_qkd2019, peng_qkd_osc2025}. Of these, we note that~\cite{wang_qkd2017} employs a slightly narrower C\nobreakdash-band bandwidth than current industry standards,~\cite{beppu_qkd2025} transmits a C\nobreakdash-band quantum signal with a fully-loaded O\nobreakdash-band classical signal, and~\cite{honz_hollowcore_qkd2023,dou_qkd2024, eriksson_qkd2019} utilize intraband coexistence, meaning that some of the available classical bandwidth must be removed for quantum use. Although it doesn't employ fully-loaded C\nobreakdash-band classical signals, previous work in our lab provides an important basis for coexistence of high-power, independent classical communications traffic with O\nobreakdash-band polarization entanglement-based communications. We have demonstrated entanglement distribution over a 48\nobreakdash-km deployed fiber loop with 18.1\nobreakdash-dBm aggregate launch power from an 11\nobreakdash-channel C\nobreakdash-band WDM source spanning 1549\nobreakdash-1565\,nm~\cite{jordan_ent_dist} and quantum teleportation over a 30\nobreakdash-km spooled fiber with 18.7\,dBm of 400\nobreakdash-Gbps classical traffic at 1547.32\,nm~\cite{jordan_tele}, proving compatibility of entanglement-based communications at coexistence power levels on the same order of magnitude as the highest demonstrated with WCS~\cite{mao_qkd2018, peng_qkd_osc2025}.

Additionally, progressing from laboratory-based experiments or demonstrations over installed fiber in a loopback configuration to deployed fiber links with separated measurement nodes is crucial for proving quantum networks are compatible with real-world classical networks. In order to enable real-time measurement between remote locations, high-precision timing synchronization is pivotal for high\nobreakdash-rate quantum sources with narrow coincidence windows. Previous works have investigated the distribution of quantum entanglement in dark fiber with GPS\nobreakdash-based synchronization, but were limited by relative timing drifts of~13\,ps/s between clocks~\cite{neumann_dark_entdist2022}. Implementations of optical synchronization over fiber have enabled near picosecond-level timing precision~\cite{argonne_entdist2025, nist_ent_dist} for quantum communication links between separated buildings within the same facility. White Rabbit synchronization, which relies on a protocol combining Synchronous Ethernet (SyncE) with IEEE-1588 Precision Time Protocol (PTP), is emerging as a leading synchronization method for high\nobreakdash-precision applications in quantum networks, with an experimentally-demonstrated timing deviation of only~4\,ps~\cite{WR_protocol,nist_conference}. Furthermore, coexistence between separated nodes has been demonstrated with independent classical communications for WCS~\cite{mao_qkd2018, berrevoets_mdiqkd2022,dynes_cambridge_qkd, peng_qkd_osc2025}, but previous works only coexist with low-power synchronization signals for entanglement-based experiments~\cite{kapoor_fermi_argonne_sync, nist_ent_dist}.  To the best of our knowledge, no prior entanglement-based experiments have demonstrated coexistence with independent classical traffic over deployed fiber without a loopback.

Our work bridges this gap; we present the first (to our knowledge) demonstration of quantum entanglement distribution coexisting with a fully-loaded C\nobreakdash-band source, enabling entanglement-based communications under conditions mimicking a real-world commercial WDM communications link (including deployed fiber, co-location with an active telecommunications exchange facility, and synchronized, physically-separated measurement nodes). The distribution of quantum entanglement and coexisting classical communications between remote, synchronized intercity nodes represents an important step towards viable, real-world quantum networks. We further present the SpRS spectrum in the O\nobreakdash-band generated by our broadband C\nobreakdash-band source, offering insights into optimal wavelength allocation for WDM quantum networking in the O\nobreakdash-band. Our quantum signal coexists with 21.4\,dBm of 1.6\nobreakdash-Tbps C\nobreakdash-band classical communications (spanning a 4.9\nobreakdash-THz bandwidth) and an L\nobreakdash-band classical synchronization channel. This synchronization signal enables real-time, synchronous quantum measurements to picosecond-level precision between remote nodes. We maintain high entanglement fidelity after entanglement distribution over a 24.4\nobreakdash-km deployed fiber link between Evanston and Chicago, proving the potential for deployed quantum networks integrated within the classical WDM communications infrastructure. 

\vspace{-5pt}
\section{Experimental Design}
\vspace{-17pt}

\begin{figure}[ht]
\centering
\includegraphics[width=8.7cm]{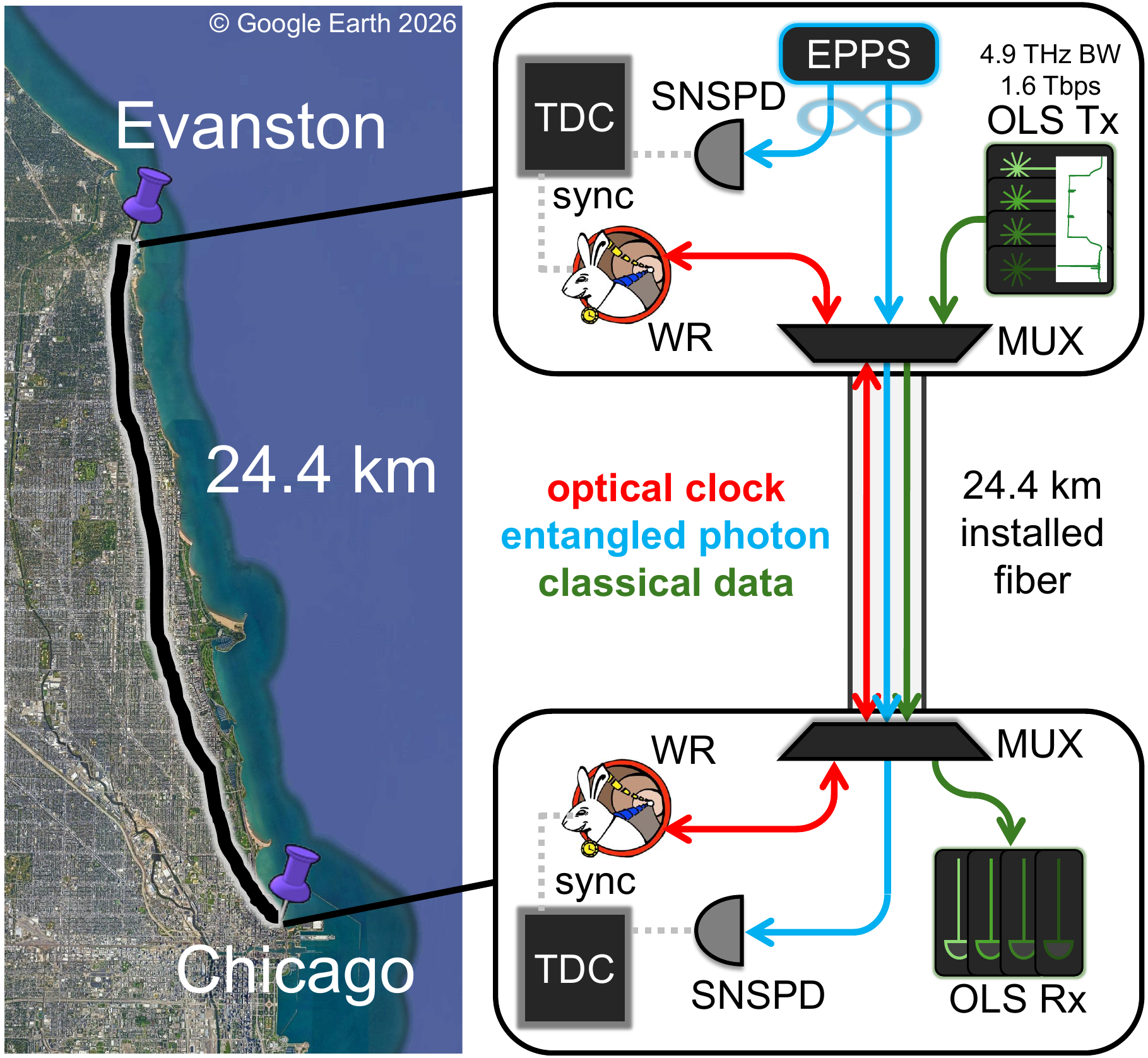}
\vspace{-7pt}
\caption{Entangled photons are distributed over 24.4\,km of deployed fiber from Evanston to Chicago. One of the photons is wavelength division multiplexed to propagate with a state-of-the-art classical communications system and optical synchronization clock. (EPPS = entangled photon pair source, OLS Tx (Rx) = classical optical line system transmitter (receiver), SNSPD = superconducting nanowire single photon detector, TDC = time-to-digital converter, WR = White Rabbit synchronizer, MUX = wavelength-division multiplexer, BW = bandwidth)}
\vspace{-8pt}
\label{exp_concept}
\end{figure}
\vspace{+2pt}
We distribute polarization-entangled photon pairs between Northwestern University in Evanston and StarLight International/National Communications Exchange Facility in Chicago, as illustrated in the conceptual diagram in Fig.~\ref{exp_concept}. We generate O\nobreakdash-band polarization-entangled photons and select a 1290\nobreakdash-nm signal and 1310\nobreakdash-nm idler through narrow spectral filtering. The 1290\nobreakdash-nm photon is distributed over a standard 24.4\nobreakdash-km installed optical fiber (SMF\nobreakdash-28) alongside 1.6\,Tbps of co-propagating state-of-the-art fully-loaded C\nobreakdash-band conventional telecommunications and a bidirectional L\nobreakdash-band synchronization signal that enables high\nobreakdash-precision synchronous photon counting between remote time-to-digital converters (TDCs). The Chicago measurement node is housed at the StarLight facility, which is a 24/7 production communications exchange and a pioneering advanced network infrastructure in Chicago, serving as a major hub for global research and education networking, integrating high-performance optical networks with software-defined networking for scientific discovery and data-intensive research~\cite{starlight}. We operate both the quantum and classical components of our system at the StarLight node under standard operating conditions for a communications exchange facility, as would be necessary for a fully-implemented quantum network within existing classical infrastructure. 

Beyond deployed network architecture, the coexistence of quantum networks with control systems such as time synchronization or channel monitoring is gaining interest because of limited fiber availability, high-precision timing measurements across separate locations, or probing perturbations to the quantum system~\cite{quantum_wrapper,bearlinQ,dcqnet_sync,nist_sync}. Here, we multiplex our quantum and classical signals with an optical clock for high-precision synchronization ($\sim 5$\,ps root-mean-square jitter) using White Rabbit (WR) timing protocols~\cite{nist_sync,WR_protocol}. This enables live, high-precision correlation measurements of entangled photon pairs between Evanston and Chicago. We further integrate our quantum signal with state-of-the-art classical transmissions; the 1290\nobreakdash-nm entangled photon is wavelength-division multiplexed with both the bidirectional L\nobreakdash-band WR clock signal and a co\nobreakdash-propagating fully-loaded C\nobreakdash-band classical system, demonstrating the potential for highly efficient use of deployed optical fiber bandwidth.

\subsection{Fully-Loaded C-band Classical Communications}
\label{subsec:OLS}
\begin{figure}[ht]
\centering
\includegraphics[width=8.6cm]{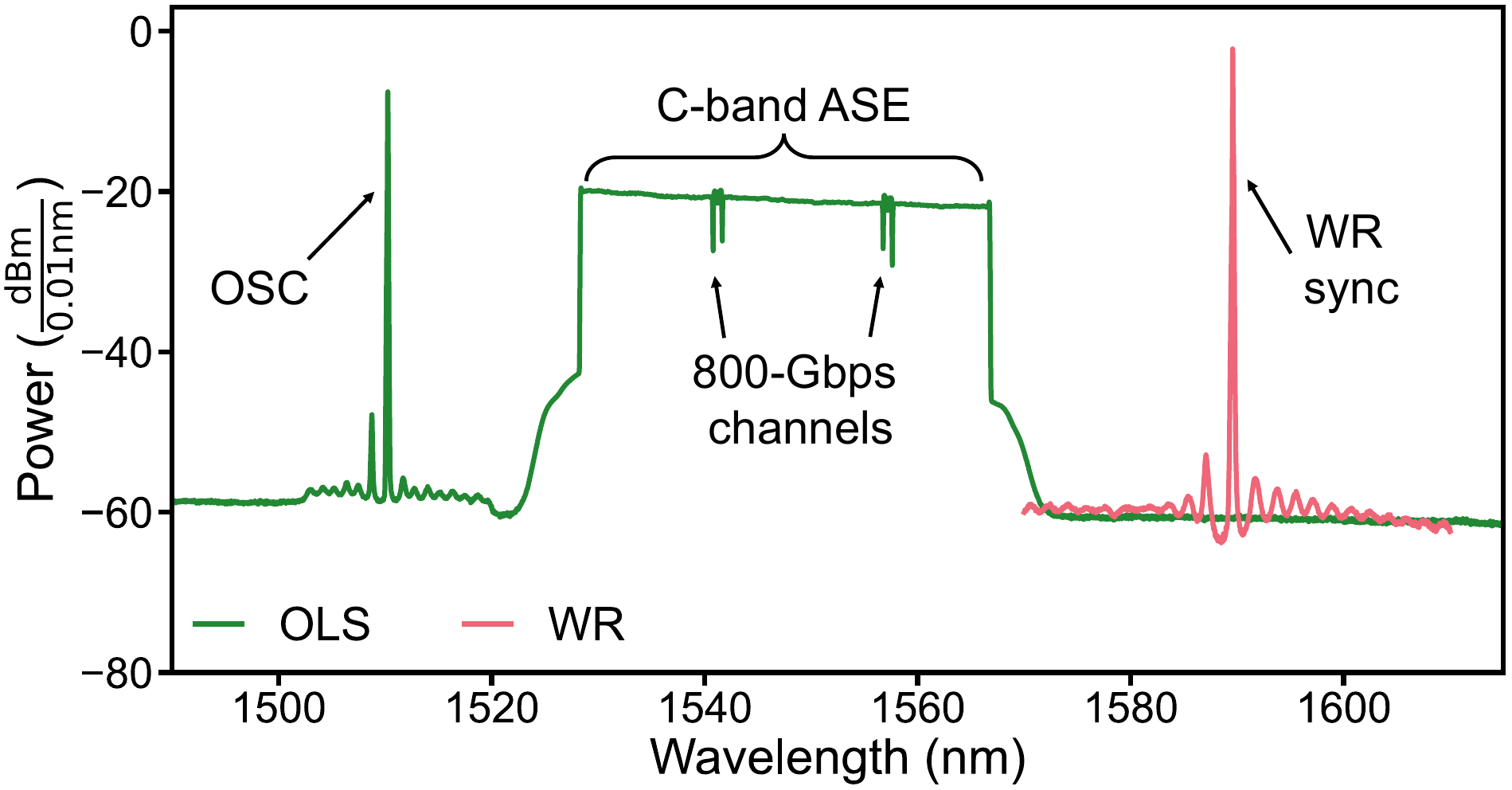}
\vspace{-5pt}
\caption{Optical spectrum analyzer (OSA) measurement of the Ciena classical OLS (green) and WR synchronization signal (red) with 0.01\,nm resolution. The peak at 1511\,nm is the OSC on the OLS and the oscillations at 1541.3\,nm and 1557.4\,nm are the 800\nobreakdash-Gbps data channels. Powers indicated here do not factor in experimental insertion losses. (WR = White Rabbit, OLS = optical line system, OSC = optical supervisory channel, ASE = amplified spontaneous emission)}
\vspace{-13pt}
\label{osa}
\end{figure}

The classical optical line system (OLS, green in Fig.~\ref{osa}), based on reconfigurable optical add-drop multiplexing technology (and comprised here of Ciena Corporation's 6500~Reconfigurable Line System, Waveserver~5, \& WaveLogic~5 Extreme), transmits an aggregate C\nobreakdash-band launch power of 21.4\,dBm spanning the 1528.4\nobreakdash-1566.9\,nm telecom wavelength range, as well as a 1511\nobreakdash-nm optical supervisory channel (OSC). The OLS also contains a 1610\nobreakdash-nm channel, which is turned on only during calibration and not depicted in Fig.~\ref{osa} as the OLS was calibrated once before performing experiments. The OLS requires two-way communication for calibration and normal operation; return communications are carried by an independent 24.4\nobreakdash-km optical fiber installed alongside the fiber used for quantum transmission and coexistence. For this experiment, the OLS is populated by two state-of-the-art 800\nobreakdash-Gbps channels centered at 1541.3\,nm and 1557.4\,nm, totaling 1.6\,Tbps. The remaining C\nobreakdash-band spectrum is automatically filled with ASE, which allows channel monitoring for system health and rapid addition of data-carrying channels to the OLS without requiring recalibration due to changing power levels~\cite{C+L_classical}. Both the data-carrying channels and ASE are amplified by an internal erbium-doped fiber amplifier calibrated such that the received power spectrum across the C-band is flat in amplitude after propagation (leading to the tilt visible in Fig.~\ref{osa}, where the optical power is measured directly after the transmitter). Because the dominant source of noise for quantum communications in this system is contributed by spontaneous Raman scattering~\cite{jordan_ent_dist}, classical power over a given wavelength span is the relevant variable for determining the amount of generated noise. Significantly, the fact that the OLS is designed to maintain the same launch power spectrum regardless of channel population means that classical data transmission in this experiment could be increased to 36.8\,Tbps without changing the impact on the quantum system. 

\begin{figure}[ht]
\centering
\includegraphics[width=8cm]{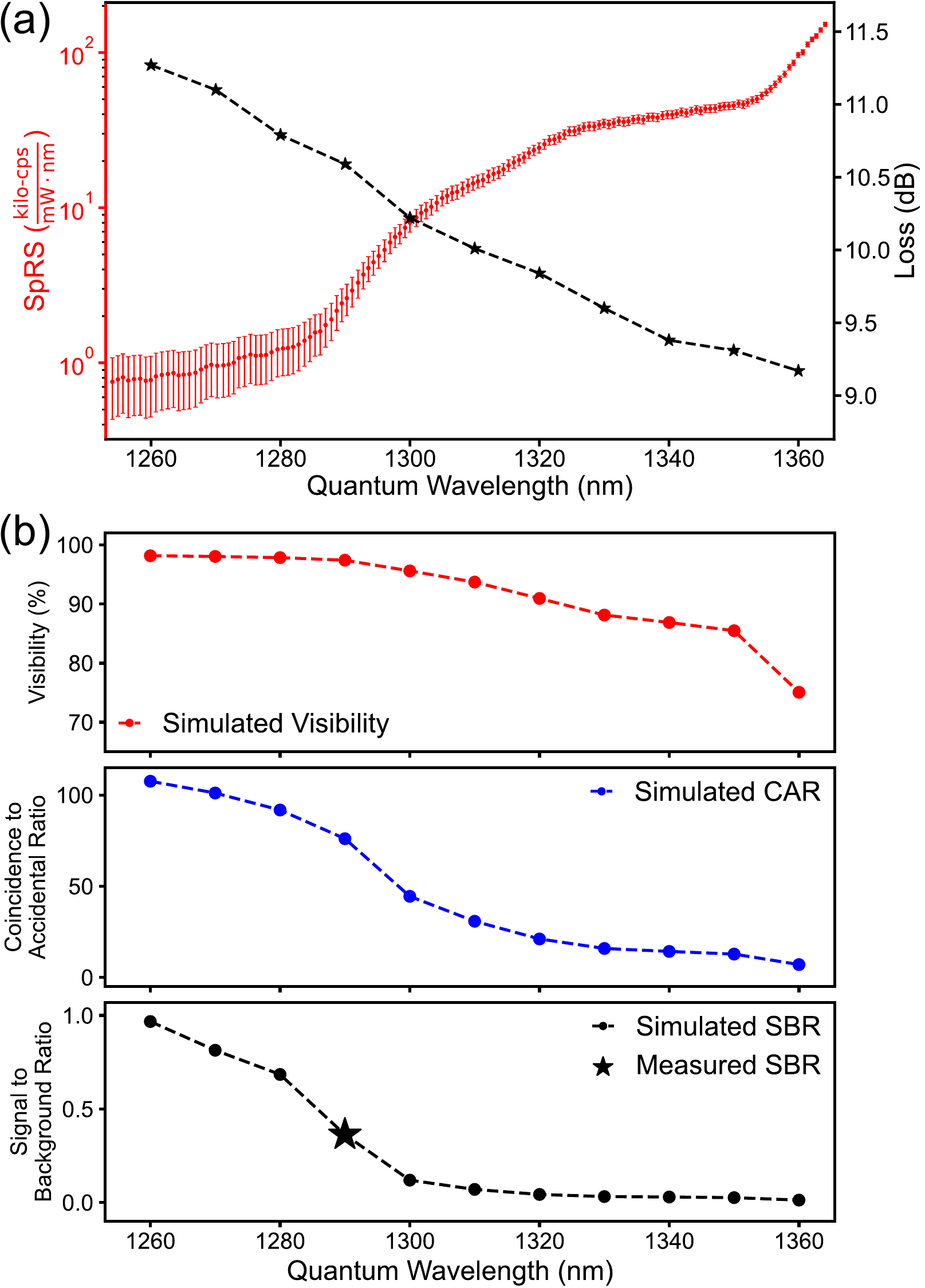}
\vspace{-5pt}
\caption{(a) Left axis (red dots): Single photon counts across the O\nobreakdash-band from spontaneous Raman scattering (SpRS) over the 24.4\nobreakdash-km deployed fiber due to the classical OLS at an 18.3\nobreakdash-dBm aggregate launch power. Counts are reported in kilo\nobreakdash-counts per second (kilo\nobreakdash-cps) normalized by classical launch power and tunable bandpass filter bandwidth. Right axis (black stars): Transmission loss over the 24.4\nobreakdash-km deployed fiber as a function of wavelength. (b) Simulations of entanglement visibility, coincidence-to-accidental ratio (CAR), and signal-to-background ratio (SBR) as a function of wavelength due to measured changes in loss and SpRS noise across the O\nobreakdash-band. Top plot (red): simulated visibility for two-photon coincidence measurements; middle plot (blue): simulated CAR; bottom plot (black): single-photon SBR simulated from the measured SBR at 1290\,nm (black star). Connecting lines are included to guide the eye.}
\vspace{-17pt}
\label{sprs}
\end{figure}

We characterize the spectrum of SpRS noise generated by the OLS as a function of wavelength across the full O\nobreakdash-band (1260\nobreakdash-1360\,nm), representing possible quantum channels. The entangled photon pair source (EPPS) used in our experiment generates a broadband spectrum due to type\nobreakdash-0 spontaneous parametric down-conversion (SPDC) that spans $\sim 40$\,nm centered at 1300\,nm. Although not employed here, this source is compatible with WDM quantum networking using multiple quantum channels for multi-node or high-rate distribution~\cite{jordan_ent_dist}. However, SpRS is not uniform across all O\nobreakdash-band channels; it has a complex spectrum generating higher or lower noise for different channels~\cite{jordan_ent_dist}. As such, it is worthwhile to study the spectrum of SpRS generated by coexisting classical sources to examine the impact of hybrid quantum-classical WDM networks. While O\nobreakdash-band SpRS spectrum characteristics from variable single-wavelength or few-wavelength sources across the C\nobreakdash-band have been studied extensively~\cite{aleksic_sprsqkd2015, sprs_model,jordan_ent_dist,peng_qkd_osc2025,wang_qkd2017}, broadband sources can create an aggregate SpRS spectrum with new features and insights for WDM quantum networking. The fully-loaded C\nobreakdash-band OLS used here represents an important case-study for broadband SpRS spectra.

The measured SpRS spectrum due to the OLS over the 24.4\nobreakdash-km fiber link is presented in Fig.~\ref{sprs}(a) alongside a plot of the transmission loss across the O\nobreakdash-band. The observed spectrum can be considered a summation of the SpRS that would have been generated by independent, single-wavelength sources across the C\nobreakdash-band. We used an 18\nobreakdash-GHz full width at half max (FWHM) tunable wavelength bandpass filter to characterize SpRS count rates from the OLS in a superconducting nanowire single photon detector (SNSPD) at the end of the 24.4\nobreakdash-km fiber with an efficiency~$\approx82\%$ across the O\nobreakdash-band. The OLS was set to a launch power of 18.3\,dBm for this measurement. At the end of the fiber, we filter out the C\nobreakdash-band light with two cascaded O\nobreakdash-band/C\nobreakdash-band wavelength-division multiplexers (WDMs), leaving only the SpRS photons and detector dark counts. The impact of dark counts (an average of 470~counts per second) is subtracted out from raw count rate measurements to focus solely on the noise counts from SpRS. Figure~\ref{sprs}(a) displays SpRS counts in log scale, normalized by launch power and filter bandwidth in kilo-counts per second (kilo-cps). 

O\nobreakdash-band quantum channels benefit from an anti-Stokes frequency detuning and reduced Raman gain generated by C\nobreakdash-band wavelengths, particularly for channels~${<1300}$\,nm~\cite{jordan_ent_dist}. For the broadband source (Fig~\ref{sprs}(a)), the spectrum of scattering noise differs from one generated by single-frequency pumps~\cite{jordan_ent_dist} as the SpRS spectrum is now a unique sum of spectra created by pumps spanning the C\nobreakdash-band. Particularly, we see higher noise at quantum wavelengths~${< 1300}$\,nm generated by the broadband light compared to a single-frequency signal at 1550\,nm, but still observe a~$\sim6$x reduction in noise by shifting the quantum wavelength from 1310\,nm to 1290\,nm. The improvement additionally comes at the cost of slightly increased transmission loss (Fig~\ref{sprs}(a)), but offers a worthwhile advantage in robustness. This advantage is demonstrated through simulations of the single-photon signal-to-background ratio (SBR), coincidence-to-accidental ratio (CAR), and visibility of entangled photon coincidence detection across O\nobreakdash-band wavelengths (Fig.~\ref{sprs}(b)). The SBR across the O\nobreakdash-band is calculated by normalizing measured loss and SpRS noise spectra to the measured single-photon SBR at 1290\,nm in our experiment (shown as a black star in Fig.~\ref{sprs}(b)), providing insights into the single channel quantum noise levels given our experimental parameters. This simulation reflects transmission over the 24.4\nobreakdash-km link with an aggregate C\nobreakdash-band classical launch power of 21.4\,dBm, as used during the coexistence experiment. 

We further simulate entanglement visibility across the O\nobreakdash-band, shown in red on the top plot of Fig.~\ref{sprs}(b). We use ${V(\lambda)=\frac{C_{\text{max}}(\lambda) - C_{\text{min}}(\lambda)}{C_{\text{max}}(\lambda)+C_{\text{min}}(\lambda)}}$ for the visibility~$V(\lambda)$ where {$C_{\text{max}}(\lambda)$ ($C_{\text{min}}(\lambda)$) is the maximum (minimum) coincidence count rate per pulse as a function of quantum wavelength~$\lambda$. The coincidence rates are calculated as~\cite{spdc_multipair}:
\vspace{-3pt}
\begin{align}
&C_{\text{max}}(\lambda) = \mu \eta_i(\lambda) \eta_s(\lambda) + S_i(\lambda) S_s(\lambda) \\
&C_{\text{min}}(\lambda) = S_i(\lambda) S_s(\lambda)
\label{eq:CC}
\end{align}
using the total channel efficiencies~$\eta_{i,s}(\lambda)$, single\nobreakdash-detector count probability~$S_{i,s}(\lambda)$, and mean photon pair number per pulse~$\mu$ generated by SPDC. Subscripts $_i$ and $_s$ refer to the idler and signal channels respectively. Wavelength dependence is added to channel efficiencies $\eta_{i,s}(\lambda)$ by scaling the experimentally measured efficiency at 1290\,nm~$\eta^{1290}_{i,s}$ by the fractional wavelength-dependent propagation loss~$X(\lambda)$ relative to 1290\,nm, such that $\eta_{i,s}(\lambda) = X(\lambda)\eta^{1290}_{i,s}$. We approximate the 1290\nobreakdash-nm channel efficiencies~$\eta^{1290}_{i,s}$ as the measured heralding efficiencies. Because these equations reflect an ideal case for alignment and entanglement generation, $C_{\text{min}}$ can be approximated as the accidental coincidence count rate per pulse. As such, we simulate the CAR using these same definitions and the relationship CAR~$=C_{\text{max}}/C_{\text{min}}$. For simulation as a function of wavelength, the singles counts $S_{i,s}(\lambda)$ are modeled according to Eq.~\ref{eq:singles}~\cite{jordan_filter, spdc_multipair}, with dark counts~$N^{\text{dark}}_{i,s}$ and SpRS noise counts~$N^{\text{SpRS}}_{i,s}(\lambda)$. Here, both $N^{\text{SpRS}}_{i,s}(\lambda)$ and $N^{\text{dark}}_{i,s}$ are the count rates per pulse, accounting for temporal filtering from a 300\nobreakdash-ps coincidence detection window.
\vspace{-5pt}
\begin{align}
S_{i,s}(\lambda) = \mu \eta_{i,s}(\lambda) + N^{\text{dark}}_{i,s} + N^{\text{SpRS}}_{i,s}(\lambda)
\label{eq:singles}
\end{align} 

We note that SpRS noise is proportional to the spectral and temporal filtering widths~\cite{jordan_filter}. Although a significant amount of noise is generated across the full spectrum by the high-power OLS, narrow spectral filtering -- particularly in the lower noise region of the O\nobreakdash-band (${< 1300}$\,nm) -- can reduce the background noise to the order of dark counts. Spectral filtering and polarization independence of noise counts are automatically factored into the simulation by using measured values in the experimental setup at 1290\,nm for $N^{\text{dark}}_{i,s}$ and $N^{\text{SpRS}}_{i,s}(\text{1290\,nm})$. In our experiment, quantum channels are filtered by 7\nobreakdash-GHz FWHM Fabry-P\'{e}rot etalons and isolated to a single etalon transmission peak using dense wavelength-division multiplexer (DWDM) filters with 100\nobreakdash-GHz channel spacing to select phase-matched photon pairs centered around the 1300\nobreakdash-nm pump for the signal (1290\,nm) and idler (1310\,nm). We filter narrowly into the joint spectrum to reject most generated SpRS (strongly limiting the effect of SpRS noise to degrade quantum transmissions) but widely enough to prevent significant reduction to the loss-independent coincidence probability \cite{jordan_filter}. The wavelength-dependent values for $N^{\text{SpRS}}_{i,s}(\lambda)$ are obtained by scaling the SpRS spectrum in Fig.~\ref{sprs}(a) to the measured SpRS counts at 1290\,nm, which further adjusts the reported spectrum to a 21.4\nobreakdash-dBm launch power and relevant insertion losses that are used to generate the results presented in Section~\ref{subsec:results} of this paper. We assume constant pair rates and detector efficiency across the O\nobreakdash-band (leading to a constant~$\mu$ and~$N^{\text{dark}}_{i,s}$). Similarly, we take propagation loss in the idler path to be equal for all wavelengths as the fiber distance is negligible (a few meters on the optical table) and we set $N^{\text{SpRS}}_i = 0$ since there is only classical light in the signal path. This model can be extended to consider configurations with coexisting classical light in both quantum paths by adding a non-zero value for $N^{\text{SpRS}}_i$. For the simulations presented in Fig.~\ref{sprs}(b), we use $\mu=0.009$, $\eta_i = 0.03$, $\eta_s = 0.001$, $N^{\text{dark}}_i = 1.8 \times 10^{-7}$, and $N^{\text{dark}}_s = 3.8 \times 10^{-7}$ as measured at 1290\,nm. 

Interestingly, the simulated CAR reduces slower with increasing quantum wavelength than the simulated SBR, which drops sharply at 1280\,nm (Fig.~\ref{sprs}(b)). The increase in tolerance to noise (also seen strongly in the visibility simulation) is due to coincidence-based detection, which helps filter vacuum emission and can lead to a higher noise tolerance for coincidence-based photon counting systems~\cite{herald, jordan_filter}. We calculate a relatively flat, high visibility for wavelengths $< 1300$\,nm. Meanwhile, for channels $> 1300$\,nm, SpRS noise increases at a higher rate than propagation loss decreases, leading to the significant drop-off in visibility seen in Fig.~\ref{sprs}(b). We see that wavelengths in the 1260\nobreakdash-1290\,nm range offer the highest entanglement visibility under coexistence with a fully-loaded C\nobreakdash-band source. Although higher loss doesn't prohibit higher rate operation (as per~\cite{jordan_ent_dist}, $\mu$ can be increased to raise rates while still maintaining visibilities higher than those for channels with stronger SpRS noise), these simulations justify our 1290\nobreakdash-nm wavelength selection. The insights gained from Fig.~\ref{sprs} can be used to select optimal quantum channel allocation or determine how other channels would be affected in a multi-channel quantum source while coexisting with a broadband C\nobreakdash-band source. 

\subsection{Synchronization Clock}
\label{subsec:sync}
\begin{figure}[h]
\centering
\includegraphics[width=8.7cm]{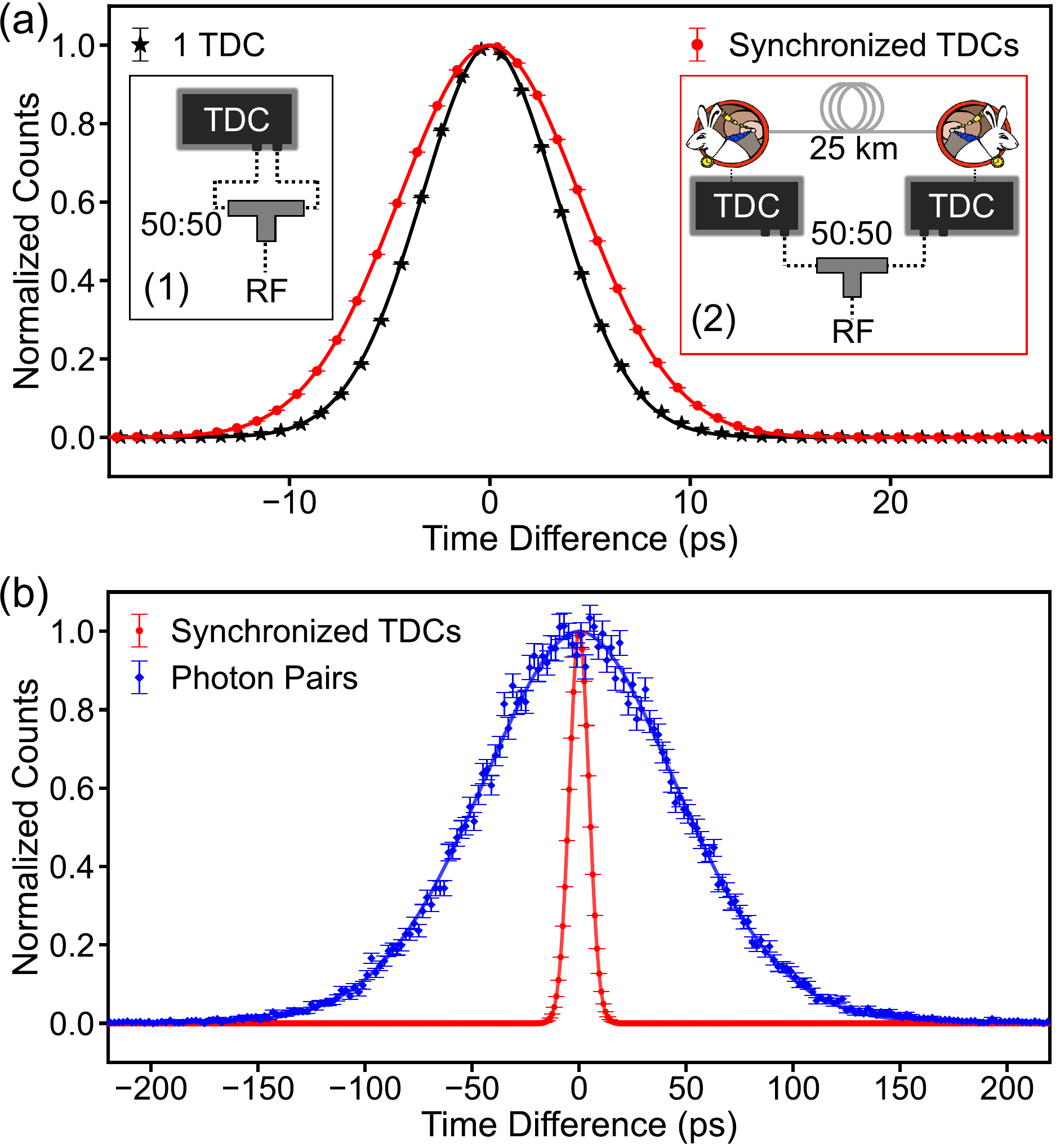}
\vspace{-5pt}
\caption{Histograms of deviation in arrival time difference. (a) Pulse arrival times relative to an identical test signal copy for either one TDC (black stars) or two TDCs synchronized by WR (red dots). Insets show diagrams of the experimental setups for jitter measurements on a 50:50 split 50\nobreakdash-MHz RF signal with (1) one TDC and (2) two synchronized TDCs. (2) is synchronized by WR with a 25\nobreakdash-km fiber spool between TDCs. Measured RMS timing jitters are $\sigma_{\text{1TDC}} = 3.5$\,ps and $\sigma_{\text{2TDC}} = 4.6$\,ps. From this, the increase in jitter due to WR synchronization is calculated to be $\sigma_{\text{WR}} = 3.0$\,ps~\cite{swabian_app_note}. (b) Comparison of timing jitter from two synchronized TDCs (red dots) relative to the coincidence peak from entangled photon pairs during the coexistence experiment over the synchronized link (blue diamonds). The RMS pulse width of the entangled photon coincidence counts is $\sigma_{\text{CC}} = 46$\,ps and coincidence windows are sized at 300\,ps to accommodate the full base of the peak. (TDC~= time-to-digital converter, RF = radio frequency, WR~=~White Rabbit, RMS = root-mean-square)}
\label{jitter}
\vspace{-15pt}
\end{figure}

\begin{figure*}[ht]
\centering
\includegraphics[width=15.6cm]{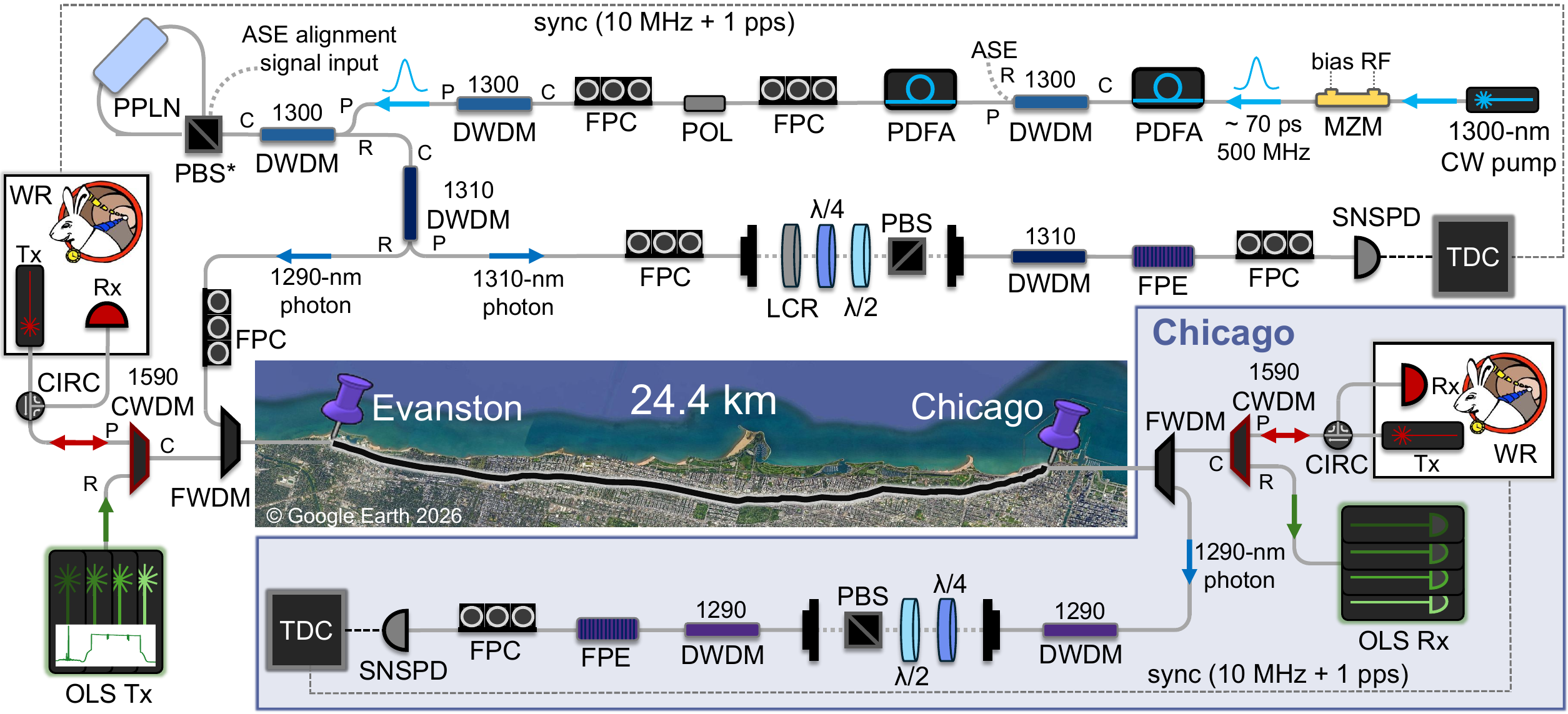}
\caption{Diagram of experimental implementation of entanglement distribution. The polarization Sagnac loop source generates polarization entangled photon pairs by second harmonic generation cascaded with non-degenerate spontaneous parametric down-conversion. We select pairs at 1290\,nm and 1310\,nm through narrow spectral filtering. The 1310\nobreakdash-nm photon is kept in Evanston, where it is filtered to ensure isolation from the pump and its polarization is characterized before it is detected. The 1290\nobreakdash-nm photon is multiplexed with the classical OLS and synchronization clock and then distributed over a 24.4\nobreakdash-km deployed fiber to Chicago. The classical signals are de-multiplexed and the 1290\nobreakdash-nm photon is narrowly filtered to remove induced SpRS noise before its polarization is characterized. Photon detections are analyzed for coincidences to verify the entangled state by two TDCs synchronized by the WR optical clock. (CWDM = coarse wavelength-division multiplexer (WDM), DWDM = dense WDM, FWDM = O\nobreakdash-band/C\nobreakdash-band WDM, CIRC = circulator, FPC = fiber polarization controller, PBS = polarizing beam splitter, PBS* = PBS with 180$^{\circ}$ phase flip on the V~path, $\lambda/2$ = half-waveplate, $\lambda/4$ = quarter-waveplate, LCR = liquid crystal retarder, PPLN = periodically-poled lithium niobate waveguide, SNSPD = superconducting nanowire single photon detector, FPE = Fabry-P\'{e}rot etalon, C = common port, P = pass port, R = reflect port, Tx = transmitter, Rx = receiver, OLS = classical optical line system, WR = White Rabbit synchronization, TDC = time-to-digital converter, pps = pulse per second, ASE = amplified spontaneous emission, PDFA = praseodymium-doped fiber amplifier, MZM = Mach-Zehnder modulator, POL = linear polarizer, RF = radio frequency signal, CW = continuous wave)}
\vspace{-5pt}
\label{exp_detail}
\end{figure*}

Following similar considerations, we select an L\nobreakdash-band synchronization channel at 1590\,nm (shown in red in Fig.~\ref{osa}), chosen for both wide wavelength separation from the quantum channel and bandwidth availability between the occupied C\nobreakdash-band spectrum and calibration channel at 1610\nobreakdash-nm for the OLS. Because Raman noise significantly decreases with increased quantum-classical channel separation, selecting an L\nobreakdash-band clock channel maintains the availability of the C\nobreakdash-band for classical data traffic while minimizing the clock's contribution to SpRS noise (much like the O\nobreakdash-band/L\nobreakdash-band wavelength allocations in~\cite{jordan_ent_dist}). Our synchronization signal is distributed bidirectionally from 1G small form-factor pluggable (SFP) transceivers (FS CWDM\nobreakdash-SFP1G\nobreakdash-ZX\nobreakdash-1590) using WR timing protocols (Safran WR\nobreakdash-LEN) to establish a common time base between TDCs (Swabian Instruments Time Tagger X) at each location with picosecond-level precision for relative time synchronization~\cite{nist_sync, WR_protocol}. WR typically operates with bidirectional SFPs at two different wavelengths~\cite{nist_sync}, but we opt to use a single wavelength to minimize occupied bandwidth; we utilize standard duplex SFPs and circulators to route signals bidirectionally over a single fiber (Fig.~\ref{exp_detail}). With the chosen large frequency detuning from the quantum channel (L\nobreakdash-band to O\nobreakdash-band) and low power requirements for propagation ($-28$\,dBm SFP receiver sensitivity), we observe a negligible increase in SpRS noise when including the bidirectional 1590\nobreakdash-nm clock signal. Due to the negligible noise even at 0\,dBm launch power, we do not fully optimize (minimize) SFP launch power; the 1590\nobreakdash-nm launch powers were $-9.2$\,dBm co-propagating and $-0.7$\,dBm counter-propagating (not to scale in Fig.~\ref{osa}; depicted measurement does not include relevant insertion losses). The co-propagating transmitter passes through three cascaded C\nobreakdash-band WDMs to remove ASE. For integration with more advanced classical systems that span the full C\nobreakdash- and L\nobreakdash-bands, the optical clock could be moved to share the O\nobreakdash-band with the quantum signal, at the cost of higher SpRS noise. However, with a sufficiently sensitive receiver (easily available off the shelf), the synchronization channel can likely be attenuated enough to allow low-error quantum communications~\cite{nist_sync}; this is precedented by studies on O\nobreakdash-band/O\nobreakdash-band coexistence for Hong-Ou-Mandel interference~\cite{me_jordan_cleo} and C\nobreakdash-band/C\nobreakdash-band entanglement distribution with powers $< -10$\,dBm~\cite{fan_entdist_and_qkd2023}. However, the channels of our OLS are restricted to the C\nobreakdash-band (plus the 1511\nobreakdash-nm OSC), allowing us to capitalize on the minimal noise configuration of placing the synchronization channel in the L\nobreakdash-band.

The WR devices send a 10\nobreakdash-MHz clock and a 1~pulse-per-second reference signal for synchronization to the TDCs, which record photon arrival times and perform correlation analysis over an internet connection using Swabian Instruments' TimeTaggerNetwork software. By placing a 300\nobreakdash-ps coincidence detection window around the entangled photon pulse arrival times, we filter SpRS noise in the time domain in addition to narrow (7\nobreakdash-GHz) spectral filtering. Figure~\ref{jitter}(a) characterizes the relative timing precision of the networked TDCs. To measure this timing jitter, we split a 50\nobreakdash-MHz radio frequency (RF) signal in half, sending either both identical signals to a single TDC (Fig.~\ref{jitter}(a), black) or one copy to each of two TDCs synchronized by WR (red). To replicate deployed conditions, we place a 25\nobreakdash-km SMF\nobreakdash-28 fiber spool in the WR clock signal path. The insets on Fig.~\ref{jitter}(a) present a diagram of these setups. We record the variation in arrival time difference between coincident signals in the histograms in Fig.~\ref{jitter}(a) with 1\nobreakdash-ps bin widths over five minutes. The widened peaks between test cases with one TDC and two TDCs indicate a 3.0\,ps root-mean-square (RMS) jitter contributed by WR synchronization (increased from $\sigma_{\text{1TDC}} = 3.5$\,ps to $\sigma_{\text{2TDC}} = 4.6$\,ps, following~\cite{swabian_app_note}). The marginally reduced performance of a single TDC compared to the specification in~\cite{swabian_app_note} can be attributed to the low trigger levels (20\,mV) used in this measurement. Figure~\ref{jitter}(b) compares the arrival time deviation histogram for synchronized TDCs with the coincidence peak of entangled photon pairs over the 24.4\,km fiber. This data was taken over 5~minutes with a 2\nobreakdash-ps bin width during coexistence with the classical OLS and includes the total timing width generated by the widths of the two photon pulses, spectral filtering, and jitter in detection and timing electronics. We find an RMS width of~46\,ps, which is an order of magnitude greater than the measured timing jitter of~4.6\,ps RMS from synchronization. For the experiment presented in this paper, this precision is more than sufficient to maintain consistent measurements within the~300-ps coincidence window (sized to accommodate the full base of the photon pair coincidence peak in Fig.~\ref{jitter}(b)). However, timing precision remains important for enabling narrow time-domain filtering of background noise and allowing use of higher repetition rate quantum sources. Even more critical will be synchronizing photon arrival times in future network applications that rely on Hong-Ou-Mandel interference, where even a few picoseconds of timing jitter begin to reduce photon indistinguishability~\cite{nist_sync}.

\subsection{Entanglement Generation and Distribution}
Figure~\ref{exp_detail} shows the optical system in greater detail. We generate polarization-encoded entangled photons though type\nobreakdash-0 cascaded second harmonic generation SPDC (c\nobreakdash-SHG\nobreakdash-SPDC) within a polarization Sagnac loop pumped at 1300\,nm~\cite{jordan_tele, cshgspdc}. The pump pulses are generated by a 1300\nobreakdash-nm continuous-wave distributed feedback laser (Eblana EP1300\nobreakdash-0DM\nobreakdash-BO1\nobreakdash-FA) that is intensity-modulated by a lithium niobate Mach-Zehnder modulator (MZM, EOSPACE AK\nobreakdash-OK5\nobreakdash-10) driven by an amplified RF signal to generate $\sim$\,70\nobreakdash-ps temporal FWHM optical pulses at a 500\nobreakdash-MHz repetition rate. The MZM is driven by narrow RF pulses created by sending a 500\nobreakdash-MHz signal through a 50:50 splitter and recombining it at an AND gate after adding a phase delay to one arm. A bias controller (Oz Optics MBC\nobreakdash-HER\nobreakdash-PD\nobreakdash-3U\nobreakdash-0V) maximizes and stabilizes the extinction ratio of the optical pulses. The pump is amplified by two cascaded praseodymium-doped fiber amplifiers (PDFAs, FiberLabs AMP\nobreakdash-FL8611\nobreakdash-OB) with a 100\nobreakdash-GHz DWDM (AC Photonics) between them to filter out ASE from the source, which is later used as a polarization basis alignment signal spanning the quantum wavelengths. The pump passes through a linear fiber polarizer, a second 100\nobreakdash-GHz DWDM to reject ASE remaining in the quantum channels, and a fiber polarization controller (FPC) to ensure an equal split ratio of pump polarization upon entering the Sagnac loop at a custom polarizing beam splitter (PBS, Oz Optics) with a 180$^{\circ}$~phase flip in the V~path and 2\%~taps for monitoring power and inputting the alignment signal. Within the Sagnac loop, a periodically-poled lithium niobate (PPLN, AdvR Inc.) waveguide is pumped bidirectionally for c\nobreakdash-SHG\nobreakdash-SPDC; the pump is up-converted to 650\,nm before down-converting inside the same PPLN waveguide to produce a 40\nobreakdash-nm wide non-degenerate spectrum of entangled photons centered around 1300\,nm. We select phase-matched entangled pairs at 1290/1310\,nm for the signal/idler, which is important to achieve the highest signal-to-noise ratio against the SpRS noise generated in the installed fiber (see Section~2\ref{subsec:OLS} and~\cite{jordan_ent_dist}). After filtering (including the 7\nobreakdash-GHz filters down the line), we generate an approximate mean photon pair number of $\mu\approx\frac{1}{CAR} = 0.009$, where CAR is the coincidence-to-accidental ratio. The generated entangled pair is aligned to the $|\Phi^+\rangle = \frac{1}{\sqrt{2}}\left( |HH\rangle + |VV\rangle \right)$ Bell state using a two-step procedure of sending a broadband polarized ASE alignment signal that spans the O\nobreakdash-band (and is filtered to the quantum signal and idler wavelengths as it propagates through the system) and minimizing single photon counts for the H/V~basis followed by non-local coincidence-based compensation of the D/A~basis using a liquid crystal retarder and FPCs to adjust for rotations during propagation~\cite{greg_alignment, jordan_tele}. The 1310\nobreakdash-nm idler is kept locally in Evanston and sent through a quarter-waveplate, half-waveplate, and polarizing beam splitter (collectively referred to as a polarization analyzer) to enable arbitrary projective polarization measurements on the two entangled photons~\cite{kwiat_tomography}. After filtering by a DWDM and a 7\nobreakdash-GHz Fabry-P\'{e}rot etalon to reject pump wavelengths, the idler photon is detected at an SNSPD (Quantum Opus Opus One) connected to the Evanston TDC (Swabian Instruments Time Tagger X). The local SNSPD has an efficiency~$>90\%$ and dark counts~$\approx1500$~counts per second.

Meanwhile, the 1290\nobreakdash-nm signal is combined with the classical OLS and synchronization signal by an O\nobreakdash-band/C\nobreakdash-band WDM before traveling over 24.4\,km of installed SMF\nobreakdash-28 optical fiber to the StarLight measurement node in Chicago. We note that the classical OLS is calibrated to transmit at maximum output power (21.4\nobreakdash-dBm aggregate launch power), which is far above the necessary power levels for error-free operation over 25\,km; we over-amplify the classical signal to demonstrate the maximum possible effect of SpRS noise on the quantum signal. The 1290\nobreakdash-nm photon path includes $\approx3$\,dB of loss before multiplexing with the classical signals. The installed fiber has insertion losses of 10.6\,dB at 1290\,nm (0.43\,dB/km) and $\approx 6$\,dB at 1550\,nm (0.25\,dB/km), which is slightly above ideal loss specifications for SMF\nobreakdash-28 fiber. The polarization and temporal delay fluctuations of this buried fiber are characterized in~\cite{gamze_pol_drift,anirudh_timedrift} under a loopback configuration. The signal polarization remained stable for the measurement duration (a few hours) of the experiment presented here and the signal photon arrival time likely drifted $<50$~ps over five hours~\cite{gamze_pol_drift}. Because the 300\nobreakdash-ps coincidence window was sized to fully accommodate the base of the Gaussian coincidence peak, this drift did not significantly impact observed coincidence rates during the measurement periods. Realignment of bases and arrival time delays was performed manually every few hours between measurement sets (but never between comparisons of dark fiber and coexistence within the same basis). Although future work seeks to implement feedback for compensation that allows long-term, continuous operation, no compensation methods were employed during this short-term experiment.

Upon arrival at StarLight, two cascaded O\nobreakdash-band/C\nobreakdash-band WDMs (Lightel WDM\nobreakdash-15\nobreakdash-A\nobreakdash-L\nobreakdash-2) de-multiplex the signals and filter out remaining classical light with $> 100$\,dB of isolation. The 1590\nobreakdash-nm synchronization channel is further de-multiplexed and sent to the WR receiver while the C\nobreakdash-band signal is routed to a receiver on the Chicago half of the OLS. The quantum channel passes through a polarization analyzer, DWDM, and 7\nobreakdash-GHz Fabry-P\'{e}rot etalon before detection at an SNSPD with an 82\%~efficiency and dark counts~$\approx 500$~counts per second (Fig.~\ref{exp_detail}). While the OLS is switched on, an additional~$\approx 24,200$~counts per second (on average) are added to the background level due to SpRS noise, reducing the CAR from~$\approx 110$ to~$\approx75$. Coincident photon detections between the two locations are analyzed over the TimeTaggerNetwork (described in Section~2\ref{subsec:sync}) to verify entanglement. Including filters, the polarization analyzer, and other fiber components, the total insertion loss for the 1290\nobreakdash-nm photon is~$\approx18$\,dB. 

\section{Results and Discussion}
\label{subsec:results}
\begin{figure}[ht]
\centering
\vspace{-5pt}
\includegraphics[width=8.3cm]{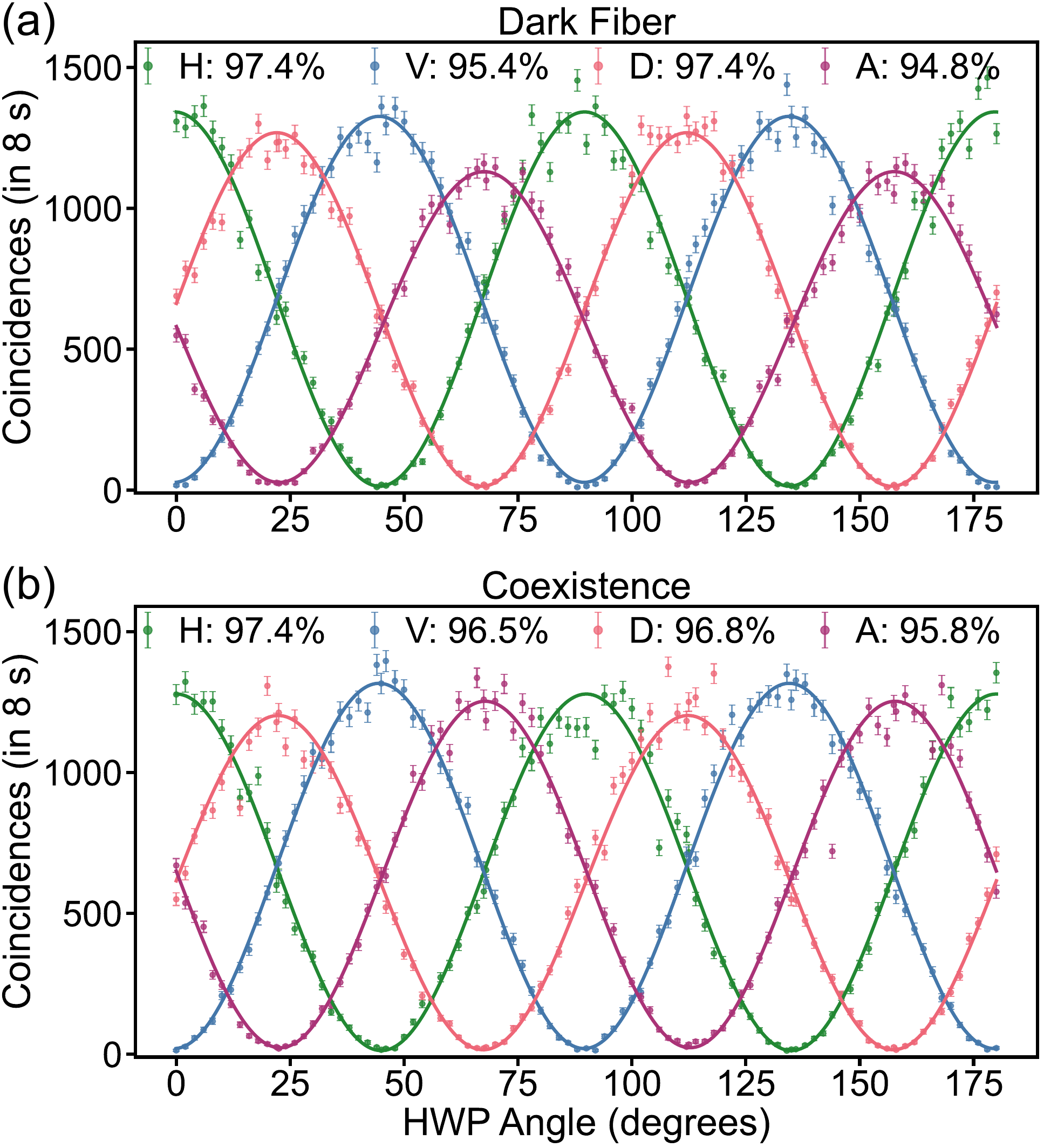}
\caption{Two-photon interference curves after entanglement distribution over the deployed 24\nobreakdash-km fiber for (a) dark fiber and (b) coexistence with the C\nobreakdash-band OLS (operating at 21.4\nobreakdash-dBm launch power) and the 1590\nobreakdash-nm optical clock. Coincidence counts are reported for 8~second intervals as a function of half-waveplate (HWP) angle for the 1310\nobreakdash-nm photon, while the polarization analyzer for the 1290\nobreakdash-nm photon is fixed in the prescribed bases. Visibilities in the H~(green), V~(blue), D~(red), and A~(purple) measurement bases are listed in the legends. All reported visibilities carry an error of $\pm0.5\%$, calculated via the Monte Carlo method and assuming Poisson statistics for photon counting.}
\label{tpi}
\end{figure}

The viability for entanglement distribution to coexist with high-power, broadband classical communications is evaluated through characterization of the two-photon interference in multiple bases and quantum state tomography after propagation over the 24.4\nobreakdash-km fiber link. Figure~\ref{tpi} presents two-photon interference curves in measured coincidences between entangled photon pulse arrivals for both dark fiber (no coexisting classical) and coexistence scenarios. In the dark fiber configuration, the 1590\nobreakdash-nm clock is routed through the independent paired fiber otherwise used for return communications to the OLS. For these measurements, the Chicago polarization analyzer is fixed in the H, V, D, or A~basis and we record coincidence counts as a function of half-waveplate angle in the Evanston polarization analyzer. We preserve $>94.8\pm0.5\%$ entanglement visibility in all measured bases ($97.4\pm0.5\%$ in the H~basis; all visibilities are presented in the legend of Fig.~\ref{tpi}) with little to no difference due to SpRS noise generated by 21.4\,dBm of coexisting classical light from the OLS and optical synchronization. It is important to note, given we use polarization-based entanglement, that SpRS noise is approximately unpolarized over long fibers~\cite{jordan_ent_dist,chapman_tomo2023}. Since noise is generated randomly across the Poincar\'{e} sphere due to polarization mode dispersion affecting the broadband classical system  over a long propagation distance, each quantum measurement basis is affected equally. Visibilities reported here are likely degraded from unity by imperfect alignment and multiphoton pair emission in the EPPS~\cite{spdc_multipair}.

\begin{figure}[ht]
\centering
\includegraphics[width=8.3cm]{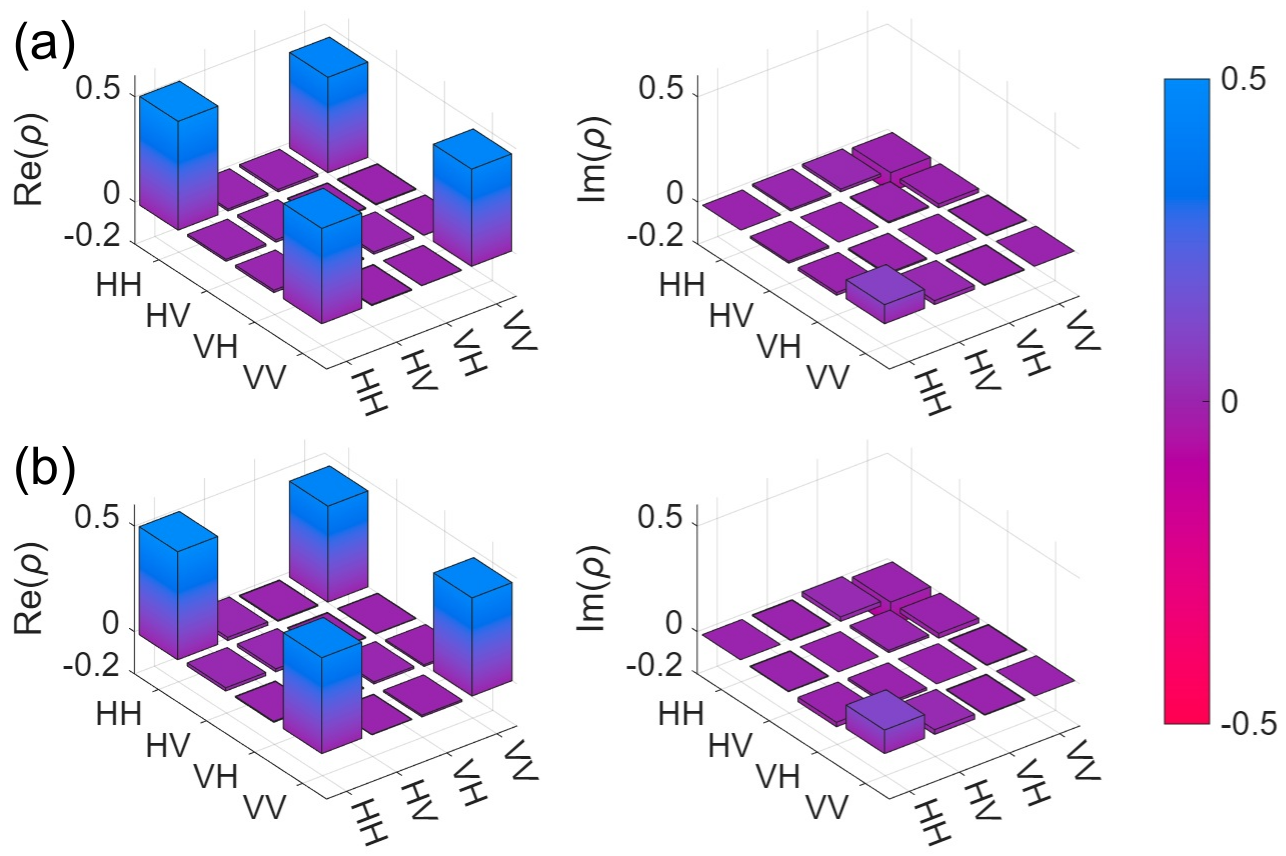}
\vspace{-5pt}
\caption{Real and imaginary parts of calculated density matrices of the entangled photon state after distribution in the 24.4\nobreakdash-km link for (a) dark fiber and (b) coexistence with the OLS and synchronization channel. Fidelity to the $|\Phi^+\rangle$ Bell state is $94.2\pm0.4\%$ for both measurements. Moreover, the coexistence scenario has a $98.8\pm0.1\%$ fidelity to the original entangled state in dark fiber. Errors are calculated via the Monte Carlo method and assuming Poisson statistics for photon counting.}
\vspace{-10pt}
\label{fid}
\end{figure}

\vspace{+5pt}
We additionally perform 2-qubit quantum state tomography on the entangled state and present the calculated density matrices in Fig.~\ref{fid}, yielding a fidelity of $94.2\pm0.4\%$ with respect to the $|\Phi^+\rangle$ Bell state for both dark fiber and coexistence scenarios. Because we observe no change in fidelity between these measurements, it is likely the reduction in fidelity from the maximally entangled $|\Phi^+\rangle$ Bell state arises from imperfect conditions, multipair emissions, and/or alignment in the entangled photon source or polarization analyzers (as is also the case for the two-photon interference measurements). As such, it is valuable to evaluate the fidelity of the entangled photon state under coexistence conditions with respect to the "original" state in dark fiber so as to more clearly estimate the impact of SpRS noise. For this analysis, we obtain a fidelity of $98.8\pm0.1\%$ to the dark fiber quantum state, indicating a minimal impact on fidelity from coexistence with the synchronization signal and high-power OLS. Errors on all measurements were calculated through Monte Carlo methods assuming Poisson statistics for photon counts for all visibilities and fidelities in Fig.~\ref{tpi} and Fig.~\ref{fid}. All reported visibilities carry an error of $\pm0.5\%$. The reported variations in visibility and fidelity between measurement scenarios are within expected errors but could be due to experimental fluctuations in the pump power or alignment of the entangled photon source over the course of data collection.

\section{Outlook}
These results demonstrate that quantum-classical networks could be implemented with little to no degradation in the quality of quantum entanglement distribution. To the best of our knowledge, this work is the first demonstration of an entanglement-based hybrid quantum-classical network alongside fully-loaded C\nobreakdash-band classical optical communications. The classical bandwidth is the widest experimentally demonstrated for entanglement-based coexistence, with classical launch powers comparable to the highest reported in previous studies on coexistence~\cite{mao_qkd2018, peng_qkd_osc2025,geng_qkd2021_ull}. Although the work presented here uses a shorter fiber distance, our previous work strongly indicates such high-power operation is feasible over longer distances~\cite{jordan_ent_dist}. Here, we used the maximum possible launch power of the commercial OLS, meaning real-world systems would be unlikely to experience stronger SpRS at the selected wavelengths. Our quantum signal co-propagated with 1.6\,Tbps of classical data, but this could easily be upgraded to over 36\,Tbps without increasing the launch power or occupied bandwidth of the classical system by replacing the auto-filled ASE with data transmission. This would have no impact on the quantum channel given that, for WDM networks, SpRS is dependent on the wavelength and power rather than the classical data rate~\cite{jordan_ent_dist, wang_qkd2017}. While this work is extremely promising, further experimentation on longer fiber links ($\sim50$\nobreakdash-100\,km) would cement the effectiveness of this technology for larger metropolitan or intercity quantum communications. Because this study benefits from decreased SpRS noise due to selecting a shorter wavelength quantum signal (O\nobreakdash-band and further shifted to 1290\,nm), long-haul communications will suffer more strongly from the increased loss compared to the C\nobreakdash-band. However, even with the additional loss, O\nobreakdash-band/C\nobreakdash-band coexistence configurations should grant higher error-free transmission rates than possible for C\nobreakdash-band/C\nobreakdash-band coexistence due to the orders of magnitude lower SpRS noise. We further provide insights on the optimal wavelength allocation for quantum channels on the O\nobreakdash-band, as well as to how a multi-channel quantum source would perform across the O\nobreakdash-band spectrum.

Moreover, this experiment provides a strong basis for future quantum-classical networks with more advanced quantum protocols. The experiment presented here, combined with prior work on coexistence for teleportation applications~\cite{jordan_tele}, suggests more advanced functions such as Bell state measurements should be attainable in similar networks. Beyond enabling shared network infrastructure with existing classical networks, coexistence could prove useful for distributing relevant information such as the Bell state measurement basis in order to fully recover the teleported state~\cite{bennett_tele}. Future work examining real-time compensation for polarization fluctuation and photon arrival time drift with integrated, coexisting classical reference signals (following similar considerations to compensation methods in~\cite{bearlinQ, gamze_pol_drift}) will help enable long-term, continuous quantum network operation.

In summary, we demonstrated O\nobreakdash-band quantum entanglement distribution with picosecond-level relative timing precision over 24.4\,km of deployed fiber alongside an L\nobreakdash-band optical synchronization clock and state-of-the-art, broadband C\nobreakdash-band telecommunications carrying 1.6\nobreakdash-Tbps classical data traffic. Through filtering and careful wavelength allocation, we maintained entanglement fidelity under the addition of high-power classical transmissions. One measurement node was housed at an active production communications exchange facility, proving compatibility with standard networking conditions. This work represents a meaningful step towards implementing quantum-classical networks under real-world metropolitan conditions.

\begin{backmatter}
\bmsection{Funding} U.S. Department of Energy (DE-AC02-07CH11359) and European Union under ERC grant QNattyNet, n.101169850.

\bmsection{Acknowledgment}
The authors would like to thank Swabian Instruments for providing the Time Tagger X units used in this experiment. We would also like to thank the members of the Advanced Quantum Networks for Scientific Discovery collaboration, led by Fermi National Accelerator Laboratory. Support for A.\,Hess was provided by the Open Quantum Initiative Undergraduate Fellowship. This work has been funded by the European Union under Horizon Europe ERC-CoG grant QNattyNet, n.101169850. Views and opinions expressed are however those of the author(s) only and do not necessarily reflect those of the European Union or the European Research Council Executive Agency. Neither the European Union nor the granting authority can be held responsible for them.

\bmsection{Disclosures}
The authors declare no conflicts of interest.

\bmsection{Data Availability}
Data may be obtained from the authors upon reasonable request.


\end{backmatter}

\bibliography{fullbib.bib}
\end{document}